# A Diagramming Technique for Teaching Students to Read Software Engineering Research Papers: an experience report

Mary Shaw
School of Computer Science
Carnegie Mellon University
Pittsburgh PA, USA
mary.shaw@cs.cmu.edu

## ABSTRACT

Reading scientific research papers is a skill that many students do not learn before entering PhD programs, but it is critical to their success. This paper describes our diagramming technique for teaching this skill, which helps them identify the structure and the scientific argument of the paper. This has made our students more effective readers.

## CCS CONCEPTS

Applied computing -> Education

## 1 Research students' heavy reading load

Conducting research in software engineering requires students to read the literature—quite a lot of literature. They must learn to read critically, selectively and effectively. This requires a different mindset from much of the reading they did as undergraduates, where they were expected to read thoroughly and, often, to accept the material as correct.

At Carnegie Mellon, first-semester PhD students in software engineering take a course that surveys software engineering research. The course has a reading load of 5-6 papers, mostly research conference papers, per week. The readings are from the primary literature, both classic and recent papers

When we originally introduced this course, students complained strenuously about the reading load. We responded by teaching them how to read the research literature selectively. Most significantly, we taught them to approach a paper not by reading it straight through but by understanding the goal of reading the paper, finding the core scientific argument of the paper, and selectively reading the parts that support their reading goal. This not only reduced the time required for the course readings, it also prepared them to approach the literature in their research projects by reading purposefully. We also now use the technique with our undergraduate research students.

This paper presents the technique we teach for finding the scientific argument in a research paper. It teaches students which elements they should expect to find and a graphical technique for locating these elements in the paper.

## 2 The scientific argument of a paper

Software engineering research seeks solutions to practical problems. This usually involves identifying a well-scoped, tractable research problem that can reasonably be expected to contribute to solving a larger practical problem.

The researcher selects a method appropriate to this idealized problem and the desired result, then conducts the research. This includes a validation step that shows how the result of the research satisfies a claim associated with the problem. There should also be an argument that the solution to the idealized problem helps to solve the real-world problem; in practice, this is often informal. Figure 1 depicts this process.

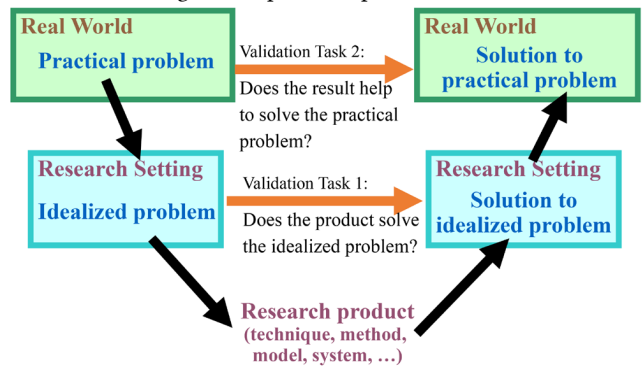

**Figure 1: Anatomy of a research result**

## 3   Helping students see the scientific argument

Many students enter graduate school accustomed to reading textbooks, in which a large body of course material is written with a single voice and uniform vocabulary, organized to present concepts in logical progression. They must approach readings from the primary literature in a very different way. Researchers write in very different voices. They use different vocabulary and formalism, and different underlying assumptions. There are overlaps or gaps between papers, and even disagreements.

The reasons for assigning papers may differ, and it's often more appropriate to read for specific points than to read for complete comprehension of the paper. Reading assignments should be clear about the reading goals of each paper, of course, but students need to translate that advice to actionable form.

Reading assignments may also include secondary literature such as reviews, surveys, essays, opinions in the form of position papers, and other material. These require their own reading skills. The primary research literature poses the greatest challenge, so that's what we address with this technique.

We teach students to approach research papers systematically. We discuss the conceptual elements of a research paper as described in [8]: the research problem/question, the kind of result, and the (coupled) research method and validation. These are abstractions that appear in most papers, realized in different ways; teaching students to look for these elements helps them to approach the papers structurally, rather than a simple linear text. For each of the conceptual elements, we discuss the different forms they may take., again based on [8]. With this background, we can approach specific papers.

First, we do a close reading of the abstract to identify the problem or research question the paper addresses, taking care to separate the real-world problem that serves as motivation from the specific, idealized research problem addressed by the paper. A critical step here is identifying the specific claim the paper makes about a new result; it is sometimes easy to confuse this with the problem that motivates the research. Then we do further close reading of the abstract to identify the research method, the new result reported in the paper, and the validation of the result—the argument the paper uses to convince the reader that the result actually solves the problem (or answers the question).

This close reading should identify the type of research the paper is reporting. In 2024, the ICSE Call for Papers [4] identified seven research areas and five review criteria that apply uniformly across research areas, but it is silent on recognizing different research paradigms. In contrast, the 2014 ICSE Call for Papers [3] asked authors to classify their research papers as analytical, empirical, technological, or methodological, and it provided different evaluation criteria for each. By identifying the type of research, we have a better idea of what to look for in the paper.[1]

---

[1] Not all abstracts are well written. If the abstract doesn't explain the paper well, the reader may have to carefully read the introduction or sometimes even more of the paper to find its structure. These cases provide teachable moments – students can learn how the quality of the abstract may affect whether the paper is read.

After the close reading we print a single sheet with thumbnails of all the pages in the paper, and we identify the text that describes the problem/question, the result, the method, and the validation. We record this by marking up the thumbnails to identify the roles that each block of text serves in the paper. The result is a one-page storyboard that shows the scientific "plot" of the paper. The thumbnails are not legible, of course, but most papers in our area have enough format variation from page to pate that it's easy to correlate the thumbnails with the version of the paper the student is actually reading.

## 4   Example

Here's how this works out on a specific example, "Two case studies of open source software development: Apache and Mozilla" by Mockus, Fielding, and Herbsleb [7]. We'll call this paper OSSD.

### 4.1   Close reading of the abstract

We begin by picking out the real-world question: investigating claims that open source software (OSS) is at least as good as commercial development methods. The "in order to" signals selection of a tractable problem, asking specifically about Apache and Mozilla (Figure 2). We can then map these to the diagram of the research process (Figure 3).

According to its proponents, open source style software development has the capacity to compete successfully … traditional commercial development methods
[In order to begin investigating such claims, we examine data from two major open source projects …]
By using [data sources] we quantify aspects of [process] and [quality] for these OSS projects.
We develop several hypotheses by comparing the Apache project with several commercial projects.
We then test and refine several of these hypotheses, based on an analysis of Mozilla data.
We conclude with thoughts about the prospects for high-performance commercial/open source process hybrids.

Real-world problem: Is OSS really at least as good as commercial?

Idealized problem: Are Apache and Mozilla as good as commercial?

Note the implicit definitions of process and quality in the lists of "aspects"

**Figure 2: Close reading of OSSD abstract for problems**

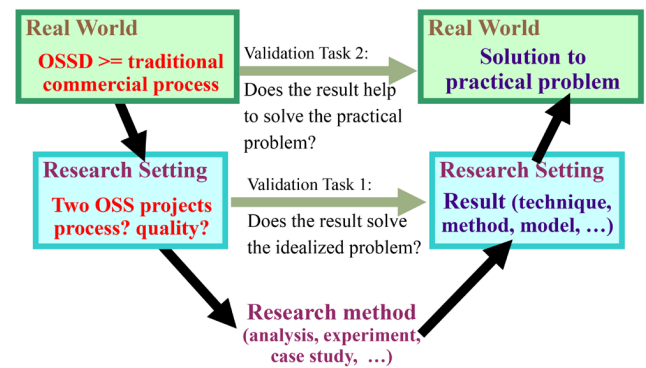

**Figure 3: Mapping OSSD problems into research plan**

We continue the close reading of the abstract by identifying the result, the research method, and the validation. The initial result is a set of hypothesis, developed with a case study of Apache. These hypotheses are validated against data from Mozilla. The research does not stop there, though: it refines the hypotheses as a second result (Figure 4). This additional analysis fleshes out the research plan (Figure 5)

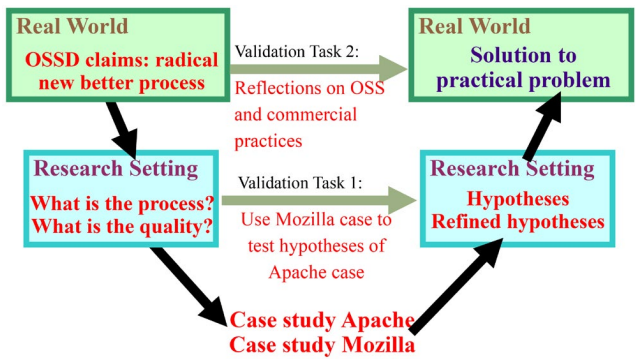

**Figure 4: Close reading of OSSD abstract for method, results, and validation**

**Figure 5: Mapping OSSD method, results, and validation into research plan**

The OSSD abstract was reasonably well written; it included everything a reader needs to get an overall sense of that the paper is accomplishing. Unfortunately, not all abstracts are this clear. In those cases it may be necessary to read (selectively!) bits of the paper to find out. The missing information is often in the introduction.

### 4.2 Finding the research design in the document

Now that we know the general plan of the paper, we can construct the storyboard of how the paper presents the research. We use colors that (approximately) match the close reading, which gives us the result is in Figure 6.

The problem statement, the baseline commercial process, and the discussion of how the work relates to the larger problem in the world are distributed through the paper (in green). The research questions and data sources follow the problem statement (in yellowish), and each case study accounts for several pages of text (in yellowish). The initial set of hypotheses lie between the case studies, as we might expect from the abstract (in orange). The initial validation and refinement of hypotheses follows, (in fuscia), and the paper concludes with the informal reflections on OSS and commercial practice (in green).

In more detail, the close reading of the abstract tells us that the paper should have two case studies and two evaluations, plus other supporting material as appropriate to case studies. The markup is aided by cues in the paper, such as topic paragraphs and section headers. It's also informed by (or perhaps helps students learn) the conventional structure of a case study paper. In this case there are two case studies, so the markup requires some engagement with the paper.

As expected, the paper begins with motivation, the question of whether OSS is at least as good as commercial software; this is marked by the green box on the second and third pages. The transition to the specific research in this paper sets up the research questions, in the yellowish box on the third and fourth pages. The close reading of the abstract established that it's a pair of case studies, so we should expect substantial sections on the two case studies. In this case the method description is combined with the case study itself; these are the two large yellowish boxes that make up the bulk of the paper, preceded by an explanation of the data sources. In the process of analysis, we see more narrative detail about the argument; for example the parallel presentation of the two case studies. A short summary of the commercial process that's the basis of comparison (in green) is tucked in between the data sources and the first case study. The initial results, the first set of hypotheses, follow the first case study and are identified in orange. The close reading showed that initial hypotheses were refined after validation, so we should expect the text that does that; it quite reasonably follows the second case study marked in fuschia. The second validation task is discharged by the discussion about the implications of the work in its expected position near the end of the paper, in green.

## 5 Experience at Carnegie Mellon

We have been using this technique to teach students to read research papers for about ten years. We originally instituted it in a first-semester PhD course in response to student complaints about the amount of time and effort required for the reading assignments. The complaints promptly subsided, and overall course evaluations improved significantly the following year (note, though, that this was only one of the changes made between the first and second course offerings).

We introduce the technique as part of the introductory lecture in this course. This lecture explains that PhD research involves a great deal of reading and that the primary research literature requires different reading skills from textbooks. We reinforce this guidance with other advice about reading papers that builds on the results of the analysis:

- *Identify what type of paper it is:* Is it a research paper, a survey, an opinion, or something else? The publication

venue often gives a hint about this. If it's a research paper, is it analytical, empirical, technological, or methodological [3]? Tailor your reading strategy to the type of paper.

- *Read with a purpose:* For class assignments, heed the guidance given in the assignment. For research reading, decide whether to skim for relevance, to read for the result, to read for the method, or to study thoroughly.
- *Read critically:* Check the reasoning—are you convinced? Evaluate the result using criteria similar to the review criteria. What's the relation between the results in this paper and other results in the field?

With this background, we introduce the diagramming technique as a "survival skill" by working through an example. We then augment the existing reading assignments for the next few lectures with the task of doing the thumbnail markup before reading the research papers.

We believe it is important for students to read the classic literature as well as current literature. The older papers are sometimes written in a style unfamiliar to current students, and expectations for the descriptions of methodology and validation have changes. The diagramming technique provides structure for navigating these older papers.

Early on some students encountered some technology issues: they did not know how to print the thumbnail pages or use a graphics editor to mark them up. We resolved this by providing instructions for printing multiple pages on a single sheet in Acrobat and being clear that markup with colored pens was fine.

Initially we only taught the technique of finding the "plot"—the storyboard on the thumbnail page. We realized that it wasn't always clear to students what elements they should look for in the paper. Adding the close reading of the abstract helped them see what they were looking for; in addition

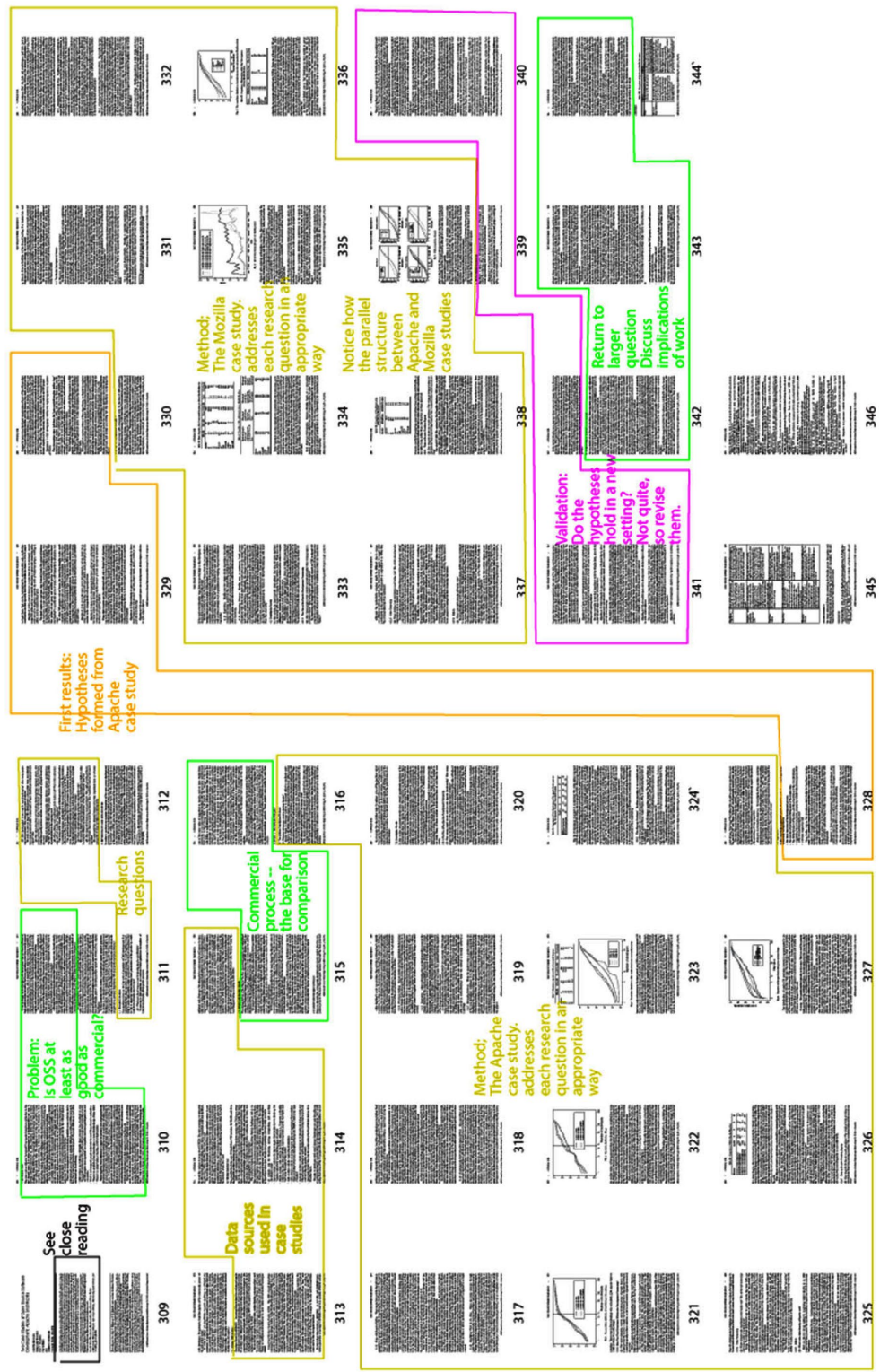

**Figure 6: Storyboard for OSSD on thumbnails. The thumbnails are not, of course, readable in this format, but in most papers the distinctive text layout on the pages helps to track the correspondence with the full-sized readable copy of the paper.**

it demonstrated the value of writing a good abstract.

Students report that after diagramming a handful of papers they get the idea—they can find the research argument without actually drawing the diagrams. This is an indication that they have internalized the skill and can take it forward to the reading required for their research projects.

More recently, we have been introducing the technique in our summer Research Experience for Undergraduates (REU) program. We introduce the technique at the beginning of the program each summer and ask the students to apply it to a research paper they're reading in connection with their summer projects. These students also respond well, and one student reported during the following academic year that he was teaching the technique to his peers.

In both cases we minimize the extra workload on students by asking them to apply the technique to material they will be reading anyhow. This also enhances the immediate relevance of the technique.

## 6   Discussion

The close reading and thumbnail annotations help students identify the overall argument and high-level organization of a research paper. Fong [1] offers advice on the full spectrum of comprehension, evaluation, and syntheses. His guidance about comprehension calls for identifying essentially the same elements as we do, but he does not take the additional step of explicitly mapping these to the text of the paper. Keshav [6] recommends reading a paper in three passes at increasing levels of detail. The first pass reads title, abstract, introduction, section headings, and conclusion. However, it does not call for a detailed close reading of the abstract to find the conceptual structure of the paper, and all three passes appear to be at the textual level, not the conceptual level.

An alternative approach would be to incorporate this in a class specifically about reading and writing research papers. In that case it could be introduced after students have already attempted to read a research paper, using that paper as an example. It could then be assigned as a task in the next several reading assignments, with the student results used as a basis for class discussion. If students write research papers in this course, they could be asked to create the storyboard before writing the text. This is a more structured assignment than the common "write an outline before you start", because it provides more guidance about what should go in the outline.

Several years before we introduced this technique in our graduate classes, I had conducted Writer's Workshops [2] in other organizations. In these workshops authors help each other improve their writing in a structured way. One of the activities calls for each participant to briefly summarize the main points of each of the other papers. They find this challenging for papers outside their immediate area, to the extent that I've been asked "Can I just use the section titles as a summary?" I believe that they would have found the summarization task less daunting if the task of finding the structure of the paper had been separated from the task of creating the summary. If I were to do this again, I would do a preliminary warmup exercise to introduce this structured approach to finding the argument of a paper.

Critical reading of the research literature is an essential skill for PhD students, as well as other students involved in research, and we are confident that this technique contributes to their proficiency.

## DATA AVAILABILITY

The annotated storyboards for seven examples are available on a Google Drive. For convenience, the drive also provides unannotated thumbnails. Because of copyright restrictions, the DOIs for the original papers are provided, not full copies. This material is available at https://drive.google.com/drive/folders/1-Li0gucyFl5W4MFylRh66BiEPon6Reyu?usp=sharing.

## ACKNOWLEDGEMENTS

This technique was developed for the PhD Core course in software engineering in the Software and Societal Systems Department at Carnegie Mellon University. It has been refined with help from faculty and students of the department. This work was supported by the Alan J. Perlis Chair of Computer Science at Carnegie Mellon University.

# Appendix: ICSE 2014 Call for Papers

The ICSE 2014 Call for Papers identified four different research paradigms (plus "perspectives") for submission. These were orthogonal to the subject options. Unfortunately, the online version of the web site for this conference is an http: site rather than an https: site, so safe web browsers may refuse to download it. Therefore the relevant bit is repeated here:

## New this year

To guide the authors in preparing their submissions and to establish a consistent set of expectations in the review process, all authors are asked, as part of the online submission process, to self-identify their papers with one or more of the following categories:

- *Analytical*: A paper in which the main contribution relies on new algorithms or mathematical theory. Examples include new bug prediction techniques, model transformations, algorithms for dynamic and static analysis, and reliability analysis. Such a contribution must be evaluated with a convincing analysis of the algorithmic details, whether through a proof, complexity analysis, or run-time analysis, among others and depending on the objectives.
- *Empirical*: A paper in which the main contribution is the empirical study of a software engineering technology or phenomenon. This includes controlled experiments, case studies, and surveys of professionals reporting qualitative or quantitative data and analysis results. Such a contribution will be judged on its study design, appropriateness and correctness of its analysis, and threats to validity. Replications are welcome.
- *Technological*: A paper in which the main contribution is of a technical nature. This includes novel tools, modeling languages, infrastructures, and other technologies. Such a contribution does not necessarily need to be evaluated with humans. However, clear arguments, backed up by evidence as appropriate, must show how and why the technology is beneficial, whether it is in automating or supporting some user task, refining our modeling capabilities, improving some key system property, etc.
- *Methodological*: A paper in which the main contribution is a coherent system of broad principles and practices to interpret or solve a problem. This includes novel requirements elicitation methods, process models, design methods, development approaches, programming paradigms, and other methodologies. The authors should provide convincing arguments, with commensurate experiences, why a new method is needed and what the benefits of the proposed method are.
- *Perspectives*: A paper in which the main contribution is a novel perspective on the field as a whole, or part thereof. This includes assessments of the current state of the art and achievements, systematic literature reviews, framing of an important problem, forward-looking thought pieces, connections to other disciplines, and historical perspectives. Such a contribution must, in a highly convincing manner, clearly articulate the vision, novelty, and potential impact.

All papers are full papers, and papers may belong to more than one category. Note that papers from any research area can fall into any of these categories, as the categories are constructed surrounding methodological approaches, not research topics (e.g., one could write an analytical paper on a new analysis technique, an empirical paper that compares a broad range of such techniques, a technological paper that makes an analysis technique practically feasible and available, or a perspectives paper that reviews the state of the art and lays out a roadmap of analysis techniques for the future).